\documentclass[conference,a4paper]{IEEEtran}
\pdfoutput=1 
\usepackage{cite}
\usepackage{amsmath,amssymb,amsfonts}
\usepackage{algorithmic}
\usepackage{graphicx}
\usepackage{textcomp}
\usepackage{xcolor}
\def\BibTeX{{\rm B\kern-.05em{\sc i\kern-.025em b}\kern-.08em
    T\kern-.1667em\lower.7ex\hbox{E}\kern-.125emX}}
    
\usepackage{gensymb}
\usepackage{subcaption}
\usepackage{balance}

\begin{document}

\title{THz Channel Sounding: Design and Validation of a High Performance Channel Sounder at 300 GHz}

\author{\IEEEauthorblockN{
Mathis Schmieder\IEEEauthorrefmark{1},
Wilhelm Keusgen\IEEEauthorrefmark{1},
Michael Peter\IEEEauthorrefmark{1}, 
Sven Wittig\IEEEauthorrefmark{1}, \\
Thomas Merkle\IEEEauthorrefmark{2},
Sandrine Wagner\IEEEauthorrefmark{2},
Michael Kuri\IEEEauthorrefmark{2},
Taro Eichler\IEEEauthorrefmark{3}
}                                     

\IEEEauthorblockA{\IEEEauthorrefmark{1}
Fraunhofer Institute for Telecommunications (HHI), Berlin, 10587, Germany, mathis.schmieder@hhi.fraunhofer.de}

\IEEEauthorblockA{\IEEEauthorrefmark{2}
Fraunhofer Institute for Applied Solid State Physics (IAF), Freiburg, 79108, Germany, thomas.merkle@iaf.fraunhofer.de}

\IEEEauthorblockA{\IEEEauthorrefmark{3}
Rohde \& Schwarz, Munich, 81671, Germany, taro.eichler@rohde-schwarz.com}
}

\maketitle

\begin{abstract}
In this paper, a novel instrument-based time-domain THz channel sounder for the 300 GHz band is introduced. The performance is characterized by conducted measurements and verified by over-the-air measurements, which confirm the sensitivity and highlight the temporal resolution of the setup. Dynamic range, maximum measurable path loss and phase stability are evaluated. The results show that the performance in terms of dynamic range and maximum measurable path loss is, to the best knowledge of the authors, unprecedented in this frequency range.
\end{abstract}


\section{Introduction}
\label{sec:introduction}

With increasing deployments of fifth generation (5G) mobile networks on a commercial scale at sub-6\,GHz and even in the millimeter wave frequencies around 28\,GHz and above, the focus in research shifts to what lays beyond 5G \cite{Latvaaho2019KeyDA}. It is expected that future wireless networks will see even further increased requirements in terms of data rate and latency. One possibility to meet the data rate requirements enabling terabit class communication is to go to even higher frequencies above 100\,GHz, where large channel bandwidths of several GHz or even tens of GHz can be used. Currently, the spectrum above 275\,GHz is not allocated and can be used for unlicensed applications with virtually unlimited bandwidth. The frequency band from 275 to 330\,GHz is therefore an attractive range for initial research in Terahertz (THz) or sub-millimeter-wave communication.

Before any communication system can be developed, the propagation characteristics of the given frequency range have to be fully understood and characterized. Only then can channel models be derived, enabling link and system level simulations of an existing or potentially new air interface standard in development for Beyond 5G. Geometry-based stochastic channel models (GSCM), like the 3rd generation partnership project (3GPP) TR 38.901 channel model \cite{3GPP-TR-38901}, valid up to 100\,GHz, are based on a large number of individual channel measurements in various scenarios. In order for the channel model to be reliable, the pool of channel measurement data has to accurately capture the specific environment. In \cite{peter2016characterization}, it was emphasized that performance metrics like dynamic range, measurable path loss, amplitude error and phase stability play a great role in the evaluation of channel sounders and in the correct interpretation of measurement data.

There are only very limited published results about channel sounders in the 300\,GHz range. In  \cite{rey2017channel}, a correlation based ultra-wideband channel sounder with an RF bandwidth of about 8\,GHz is presented. The dynamic range is approximated to be 60\,dB, but, due to lack of back-to-back calibration, the channel impulse responses (CIRs) show false additional components caused by non-linearities in the frequency extenders.

This paper presents a novel instrument-based time-domain channel sounder for the frequency range between 280 and 330\,GHz. In Section \ref{sec:setup}, the setup and key parameters are introduced followed by a brief description of the THz transceiver modules in Section \ref{sec:thz}. An evaluation of the performance in terms of dynamic range, measurable path loss and phase stability through conducted measurements is given in Section \ref{sec:evaluation}. The channel sounder was also evaluated using over-the-air (OTA) measurements, which is briefly described in Section \ref{sec:ota}.

\section{Channel Sounder Setup and Parameters}
\label{sec:setup}

The channel sounder was set up at the laboratory at Fraunhofer HHI in Berlin. It is based on advanced test and measurement equipment and specific components developed by Fraunhofer HHI and Fraunhofer IAF. It uses the principle of time domain channel sounding \cite{salous2013radio}. 

\begin{figure}[htbp]
\centerline{\includegraphics[width=0.45\textwidth]{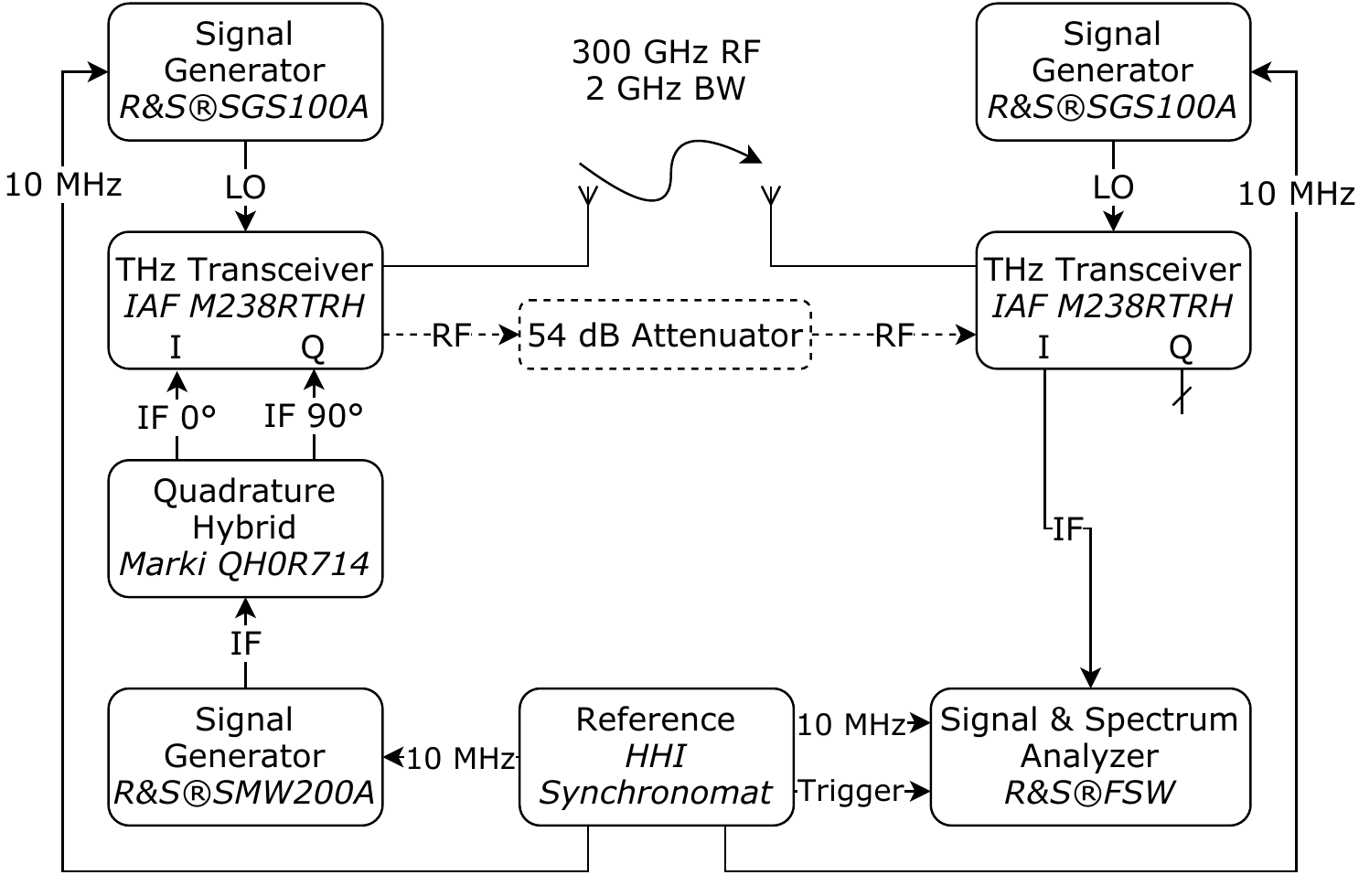}}
\caption{Block diagram of the channel sounder setup at 300\,GHz}
\label{fig:block_setup}
\end{figure}

In Fig. \ref{fig:block_setup}, a block diagram of the components used for the measurements is shown. At the transmitter (Tx), a 2\,GHz wide channel sounding sequence is generated by a vector signal generator (R\&S\textsuperscript{\textregistered}SMW200A) at an intermediate frequency (IF) of 12\,GHz with a sampling rate of 2.4 Gsamples/s at a resolution of 14\,bit. In order to achieve single sideband upconversion, a 3\,dB quadrature hybrid (Marki Microwave QH-0R714) is used to generate 0\,\textdegree{} and 90\,\textdegree{} phase shifted variants of the IF signal, which is then upconverted within the THz IQ transceiver module (M238RTRH by Fraunhofer IAF) to an RF frequency between 270 and 320\,GHz, depending on the local oscillator (LO) frequency. The THz transceiver is further described in Section \ref{sec:thz}. With this setup, the achieved suppression of the lower sideband is about 15 dB, which probably could be further increased by calibration of the IQ mismatch. A signal generator (R\&S\textsuperscript{\textregistered}SGS100A) supplies the LO signal at a frequency of 8\,GHz with a power of $-6$\,dBm which gets multiplied by 36 to 288\,GHz inside the THz transceiver module. The RF signal is available at a WM-864 waveguide at a frequency of 300\,GHz with a maximum power of $-2.2$\,dBm where it can be transmitted over the air or through a waveguide attenuator. In the current transceiver module the LO suppression is quite low, resulting in an LO leackage of around -10 dBm in the THz range.

At the receiver (Rx) side, a similar heterodyne principle as at the transmitter is used, but in a double sideband configuration. Single sideband downconversion using a quadrature hybrid is possible, but no gain in precision and accuracy is expected by the authors of this paper. In this case, the LO is generated at a frequency of 8.15\,GHz with a power of $-6$\,dBm and multiplied by 36 to 293.4\,GHz, resulting in an IF of 6.6\,GHz for the upper sideband signal which is sampled by a signal and spectrum analyzer (R\&S\textsuperscript{\textregistered}FSW) at a rate of 2.4 Gsamples/s and a resolution of 16\,bit. The baseband samples are then transferred to a PC for analysis. 

The LO and IF frequencies at transmitter and receiver are chosen in a way that no unwanted images or LO signals generated by the transmitter side appear into the upper sideband of the receiver. In general, the IF at the transmitter side should be chosen quite high, to separate LO feedthrough and lower sideband image from the signal well, whereas at the receiver side the IF frequency should be as low as possible, to ensure the best sensitivity of the 300\,GHz IQ mixers. Since the THz transceiver supports an IQ bandwidth of more than 20\,GHz, extremely wide-band channel-sounding would be possible. Theoretically a direct IQ modulation and demodulation should also be possible with the setup, which was not investigated so far.

Both transmitter and receiver share a common 10\,MHz reference provided by a high precision rubidium clock from Fraunhofer HHI which also triggers the receiver. Two Synchronomats can also be phase locked by back-to-back calibration and then disconnected. This way, the transmitter and receiver can be set up without a wired connection, allowing very flexible measurements and rapid relocation. While this has not been tested yet at 300\,GHz, results from channel sounding campaigns at other frequencies suggest no issues.

\begin{figure}[htbp]
\centerline{\includegraphics[width=0.5\textwidth]{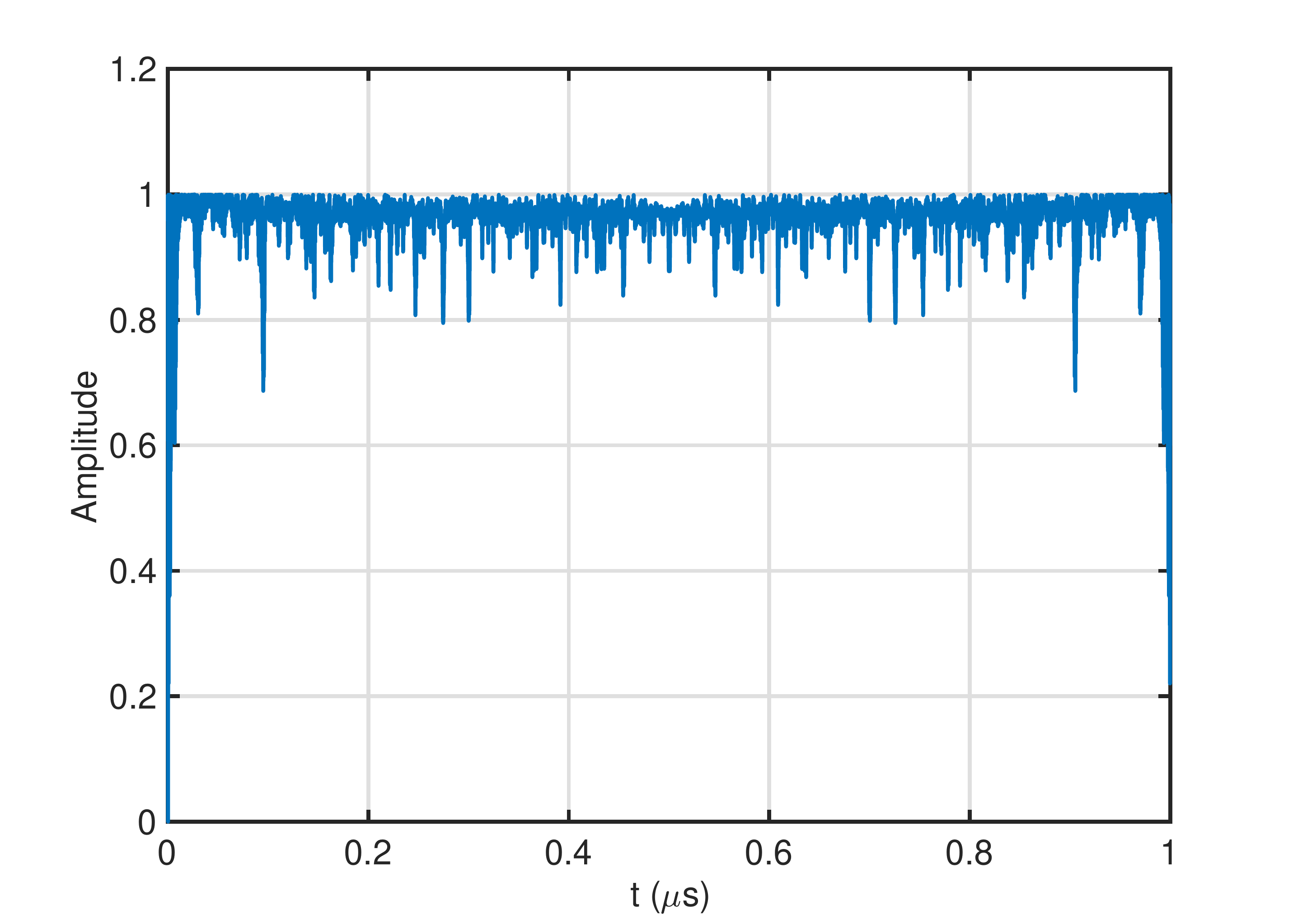}}
\caption{Magnitude of the correlation signal over time}
\label{fig:correlation}
\end{figure}

A complex periodic Frank-Zadoff-Chu (FZC) sequence \cite{friese1997multitone} was chosen as sounding signal for its optimal correlation properties. By an iterative optimization procedure, which is based on \cite{friese1997multitone}, the crest factor could reduced from $2.6$\,dB to an extremely low value of $0.3$\,dB, which allows for a low power back-off and highest transmit power.
Figure \ref{fig:correlation} shows the magnitude of the correlation signal over time, which corresponds to the normalized envelope of the signal and highlights the low crest factor.
The sequence has $N_{\mathrm{seq}} = 2000$ elements and a duration of 1\,\textmu s, resulting in a (theoretical) correlation gain for uniform sequences of $G_{\mathrm{corr}} = 33$\,dB according to \eqref{eq:correlationgain}.

\begin{equation}
    G_{\mathrm{corr|dB}} = 10 \cdot \log_{10}(N_{\mathrm{seq}})
    \label{eq:correlationgain}
\end{equation}

The length of the sequence was chosen to cope with the expected maximum channel length of 1\,\textmu s. Factoring in further gain from coherent averaging, the processing gain could be further increased by an averaging gain according to \eqref{eq:averaginggain}.

\begin{equation}
    G_{\mathrm{avrg|dB}} = 10 \cdot \log_{10}(N_{\mathrm{avrg}})
    \label{eq:averaginggain}
\end{equation}

Since the setup allows for 50\,ms snapshot length, a coherent averaging over $N_{\mathrm{avrg}} = 50,000$ sequences was implemented, resulting in additional gains of $47$\,dB from averaging. Therefore the overall theoretical processing gain \cite{peter2016characterization} calculates to $G_{\mathrm{proc|dB}} = 80$\,dB by adding up the correlation and averaging gain. 

Table \ref{tab:setup_parameters} displays the fundamental parameters and capabilities of the channel sounder. The actual sounder performance is evaluated in Section \ref{sec:evaluation}.

\begin{table}[h!]
\caption{Channel Sounder Setup and Parameter Settings}
\begin{center}
\begin{tabular}{ll}
\hline
\textbf{Parameter} & \textbf{Value and Unit} \\
\hline
Carrier frequency & 300\,GHz \\
Bandwidth $B$ & 2000\,MHz \\
Sampling rate at Tx \& Rx & 2400\,MHz \\
Maximum power at Tx port & $-2.2$\,dBm\\
Sounding sequence duration & 1\,\textmu s\\
Number of averages $N_{\mathrm{avrg}}$ & 50,000\\
Measurement time $T_\mathrm{m}$ & 50\,ms \\
Theoretical processing gain $G_{\mathrm{proc}}$ & 80\,dB \\
\hline
\end{tabular}
\label{tab:setup_parameters}
\end{center}
\vspace{-1em}
\end{table}

\subsection{Evaluation of Waveguide Components}

For evaluation of the attenuator and horn antennas which were used for the verification of the channel sounder, a vector network analyzer (R\&S\textsuperscript{\textregistered}ZVA67) was used together with a frequency extender (R\&S\textsuperscript{\textregistered}ZC330). Since no suitable WM-864 calibration kit was available, the network analyzer could only be calibrated using two-port normalization. The through calibration was achieved by directly connecting the frequency extender's waveguide flanges and isolation was accomplished by shorting one port. During the characterization measurements, a fixed attenuator was used. It showed a flat attenuation of 54\,dB over the frequency range between 270 and 320\,GHz. The OTA measurements were conducted using open 5\,cm long WM-864 waveguides and corrugated horn antennas (RPG FH-PP-330). The waveguides themselves showed an attenuation of $1.8$\,dB per 5\,cm.
In order to characterize the gain of the open waveguides and horn antennas in boresight direction, the frequency extenders were set up in a way that the antennas were exactly 1 meter apart, whereas potential reflections from the measurement table were suppressed by using absorbing material. By comparing the measured $S_{21}$ to the theoretical free-space path loss (FSPL) for 1 meter at 300\,GHz of $82$\,dB and the assumption that both waveguides and horn antennas have similar properties, the gain of the open waveguides was evaluated to be $4.8$\,dBi and of the horn antennas to be $26$\,dBi. Considering $1.8$\,dB of loss in the waveguide, the performance is similar to commercial open waveguide probe antennas with an advertised gain of $6.5$\,dBi.

\subsection{Evaluation of Transmit Power}

The channel sounder's transmit power and its relation to the IF signal power was evaluated using a VDI Erickson PM5B power meter with a WM-864 to WM-2540 taper.

\begin{figure}[hbtp]
\centerline{\includegraphics[width=0.45\textwidth]{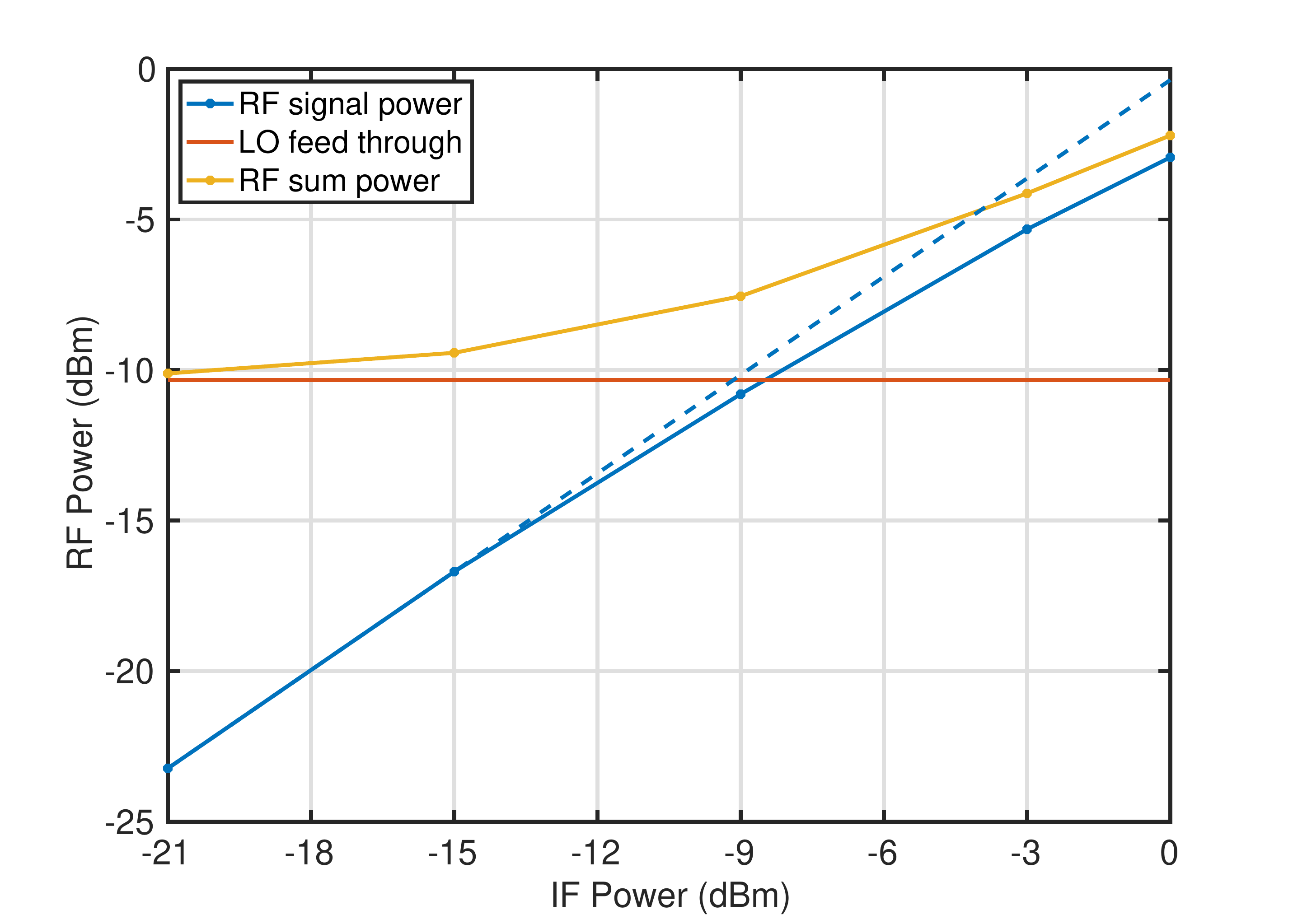}}
\caption{Transmit power in relation to IF signal power}
\label{fig:power}
\end{figure}

As illustrated in Fig. \ref{fig:power},
the LO feed through is a constant, significant part of the RF sum power with $-10.3$\,dBm. The RF signal power starts to compress slightly for IF powers above $-15$\,dBm. For the measurements, the IF power was set to $-3$\,dBm, resulting in a 1.7\,dB compressed RF signal with a power of $-5.3$\,dBm and a RF sum power of $-4.2$\,dBm. Maximum output power can be achieved with an IF power of $0$\,dBm, where the RF signal power is $-2.9$\,dBm with $2.5$\,dB of compression and the RF sum power is $-2.2$\,dBm. 

\section{THz Transceiver Frontend}
\label{sec:thz}

\begin{figure}[htbp]
\centerline{\includegraphics[width=0.45\textwidth]{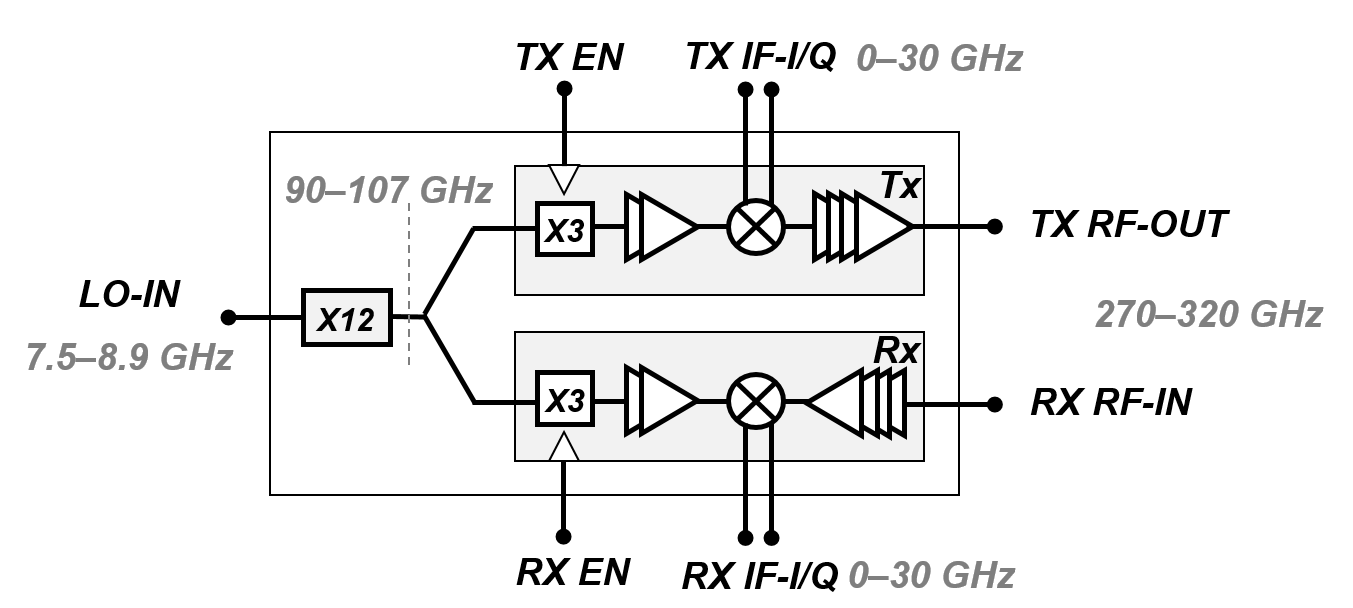}}
\caption{Functional block diagram of the 300\,GHz transceiver waveguide module. Grey components are realized as individual MMICs.}
\label{fig:thz_block}
\end{figure}

The block diagram of the 300 GHz transceiver is shown in Fig. \ref{fig:thz_block}. The module supports an external local oscillator reference signal between 7.5 and 8.9\,GHz. The signal is split after the frequency multiplier MMIC (X12) by a 3-dB waveguide coupler. The two synchronous signals are input to the Tx respectively Rx MMIC, which multiply them to a local oscillator frequency in the band from 270 to 320\,GHz. The fundamental broadband mixer uses a direct conversion architecture. The in-phase (I) and quadrature (Q) baseband signals have a bandwidth of more than 20\,GHz. The Tx MMIC integrates a broadband medium power amplifier and the Rx MMIC integrates a high gain low noise amplifier. The employed MMICs were implemented using a 35\,nm metamorphic HEMT technology with an In$_{0.52}$Al$_{0.48}$As/In$_{0.80}$Ga$_{0.20}$As channel grown on 4$''$ GaAs wafers.

\begin{figure}[htbp]
\centerline{\includegraphics[width=0.5\textwidth]{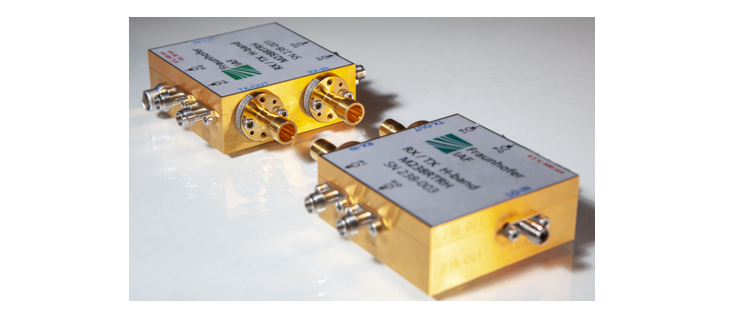}}
\caption{Photograph of the developed waveguide modules}
\label{fig:thz_module}
\end{figure}

The developed transceiver was packaged in a WM-864 waveguide module using the split-block approach (Fig. \ref{fig:thz_module}). The polarization of the standard gain horn antennas can be rotated by E/H twists, which are implemented by quarter-wavelength shims with bow-tie aperture. The size of the module is 60 x 60 x 23 mm$^3$ (LxWxH). An integrated micro-controller sets the bias conditions of the Rx, Tx and X12 MMICs. The Tx and Rx channel can be separately turned off by the enable logic signal for an optional time division operation.

\section{Conducted Evaluation Measurements}
\label{sec:evaluation}
In order to evaluate the performance of the channel sounder in terms of dynamic range, measurable path loss and phase stability, the transmitter and receiver were connected via a fixed $54$\,dB attenuator and 50,000 channel impulse response (CIR) snapshots were recorded after a back-to-back calibration. The RF signal power was set to $-5.3$\,dBm. A Chebyshef window with $80$\,dB sidelobe suppression was applied to the received signal in frequency domain.

\begin{figure}[htbp]
\centerline{\includegraphics[width=0.45\textwidth]{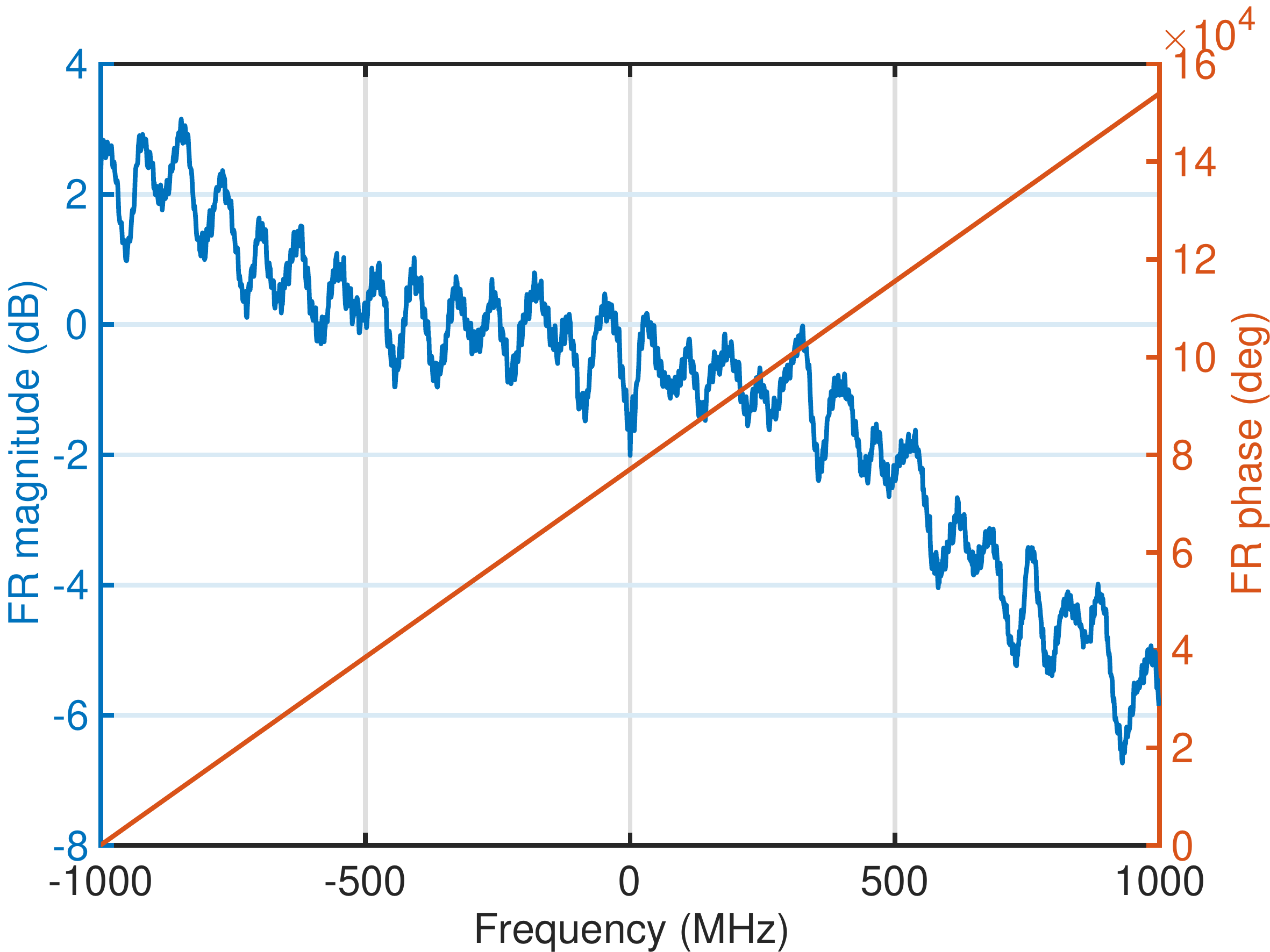}}
\caption{Back-to-back system frequency response}
\label{fig:fr_system}
\end{figure}

During back-to-back calibration, the system frequency response (FR) of the entire Tx and Rx chain was determined. Figure \ref{fig:fr_system} shows that the FR magnitude varies approximately 10\,dB across the bandwidth and the FR phase shows a strong linear increase to more than $150,000$\,\textdegree{}. The linear course of the phase represents the time offset of the trigger in the receiver in relation to the start of a period of the sounding signal. The offset is random, but fixed as long as the signal generator is not turned off.

\begin{figure}[hbp]
\centerline{\includegraphics[width=0.45\textwidth]{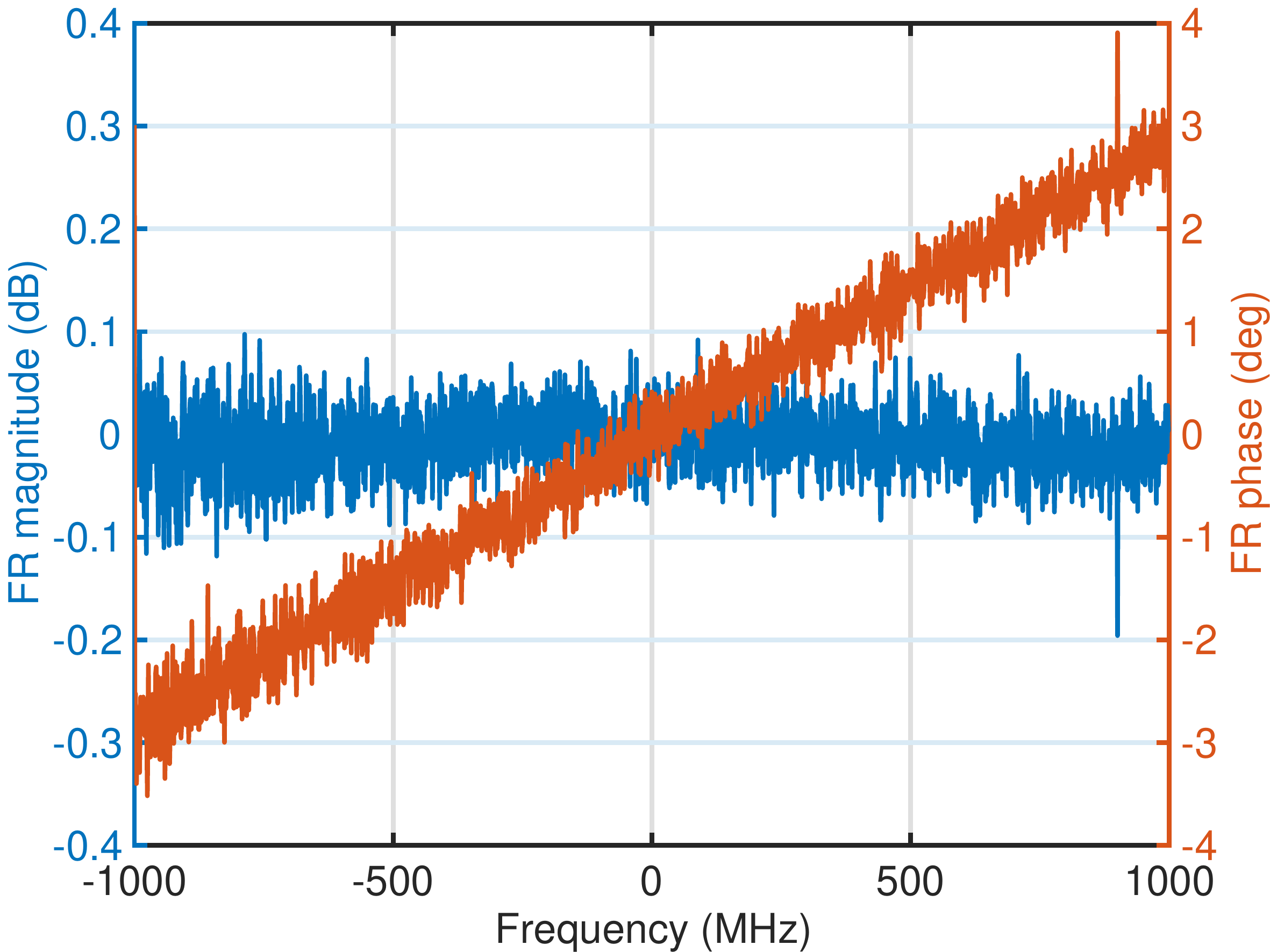}}
\caption{Calibrated system frequency response}
\label{fig:fr_system_calibrated}
\end{figure}

This back-to-back measurement was then used to calibrate the system, leading to a calibrated system frequency response as shown in Fig. \ref{fig:fr_system_calibrated}.
The FR magnitude shows only minimal variations of 0.2\,dB. A small residual, linear phase variation remains, which may be caused by trigger jitter. However, the variation is only 6\,\textdegree{}, corresponding to a temporal inaccuracy of about 8\,ps. 

\begin{figure}[htbp]
\centerline{\includegraphics[width=0.5\textwidth]{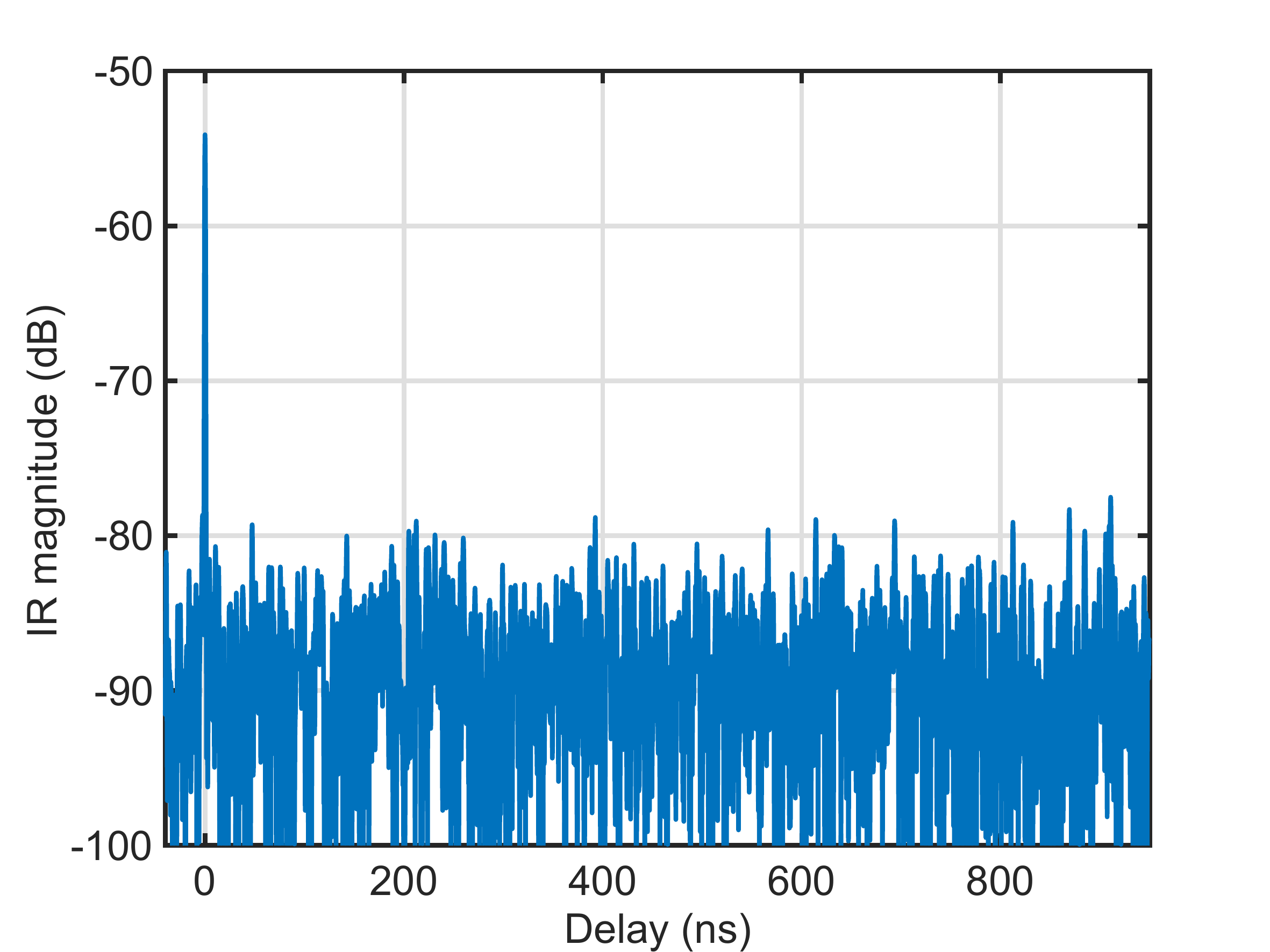}}
\caption{Single impulse response snapshot without averaging}
\label{fig:cir}
\end{figure}

In Fig. \ref{fig:cir}, a single impulse response (IR) snapshot without averaging is displayed. It shows the single main peak with a relative power of $-54$\,dB at 0\,ns delay and a mean relative noise level of $-87.1$\,dB. In order to reduce the noise level, coherent averaging over several IR snapshots can be used. By averaging over all 50,000 IR snapshots, the theoretical processing gain and therefore the maximum expectable reduction in noise level is $47$\,dB. For coherent averaging to be possible, the phase has to be stable over all IR snapshots. 

\begin{figure}[htbp]
\centerline{\includegraphics[width=0.45\textwidth]{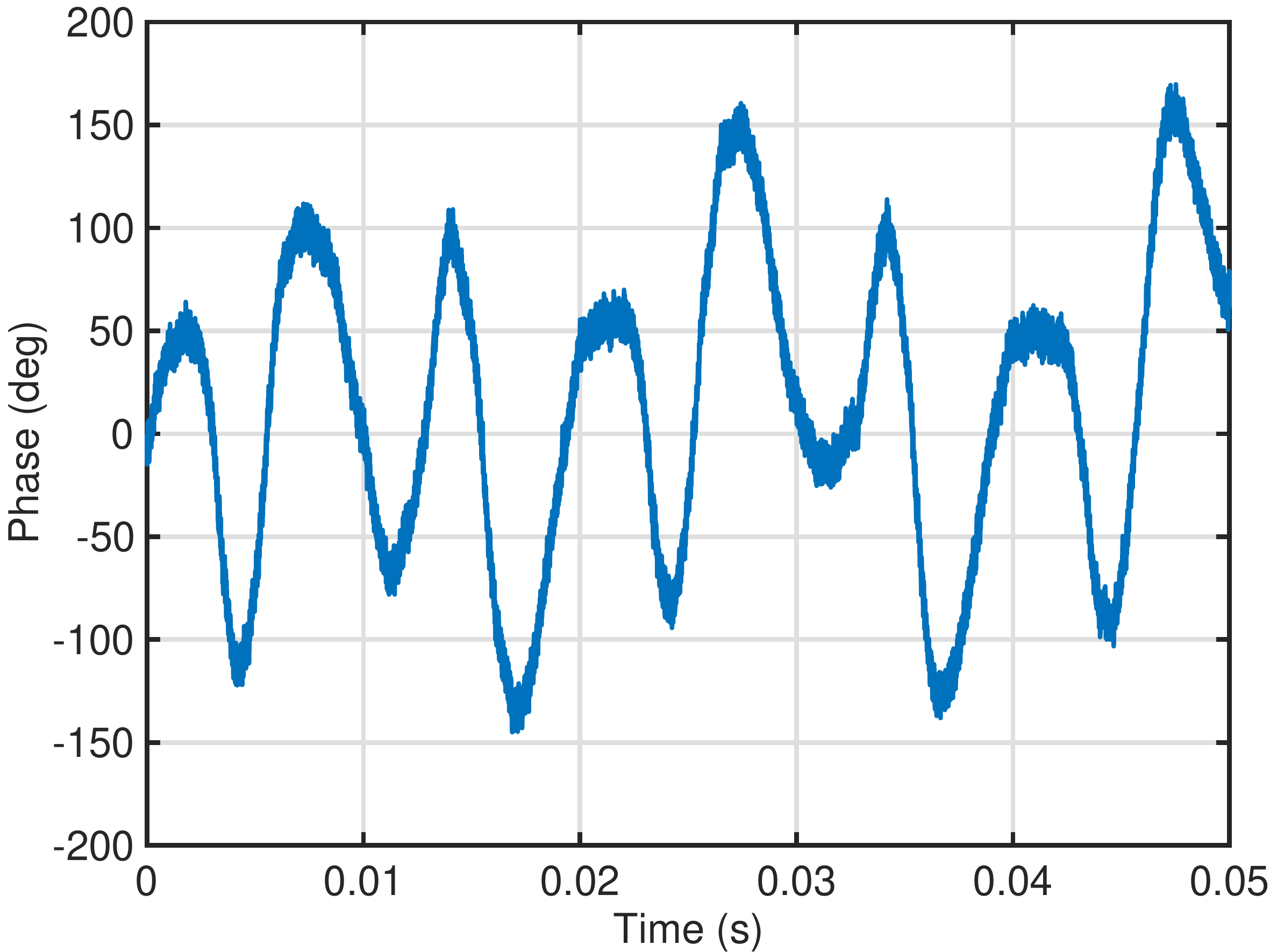}}
\caption{Evolution of the phase (IR maximum) over time}
\label{fig:phase}
\end{figure}

As shown in Fig. \ref{fig:phase}, the phase varies more than 300\,\textdegree{} over the measurement period of 50\,ms. Because the scenario is static, all variations in phase are system-intrinsic, mainly caused by the high multiplication of the reference signal. It is therefore possible to track the phase of the main peak in every IR snapshot and normalize it by multiplying with the inverted phase. This is also possible in over-the-air measurements, as long as the scenario is static and the strongest propagation path has sufficient prominence in the IR. If the signal-to-noise ratio (SNR) is too low, longer sequences with a higher intrinsic processing gain have to be used or pre-averaging over a smaller number of sequences where the phase is sufficiently stable has to be applied. It has to be noted that this averaging method was also applied to the back-to-back calibration measurement.

\begin{figure}[htbp]
\centerline{\includegraphics[width=0.5\textwidth]{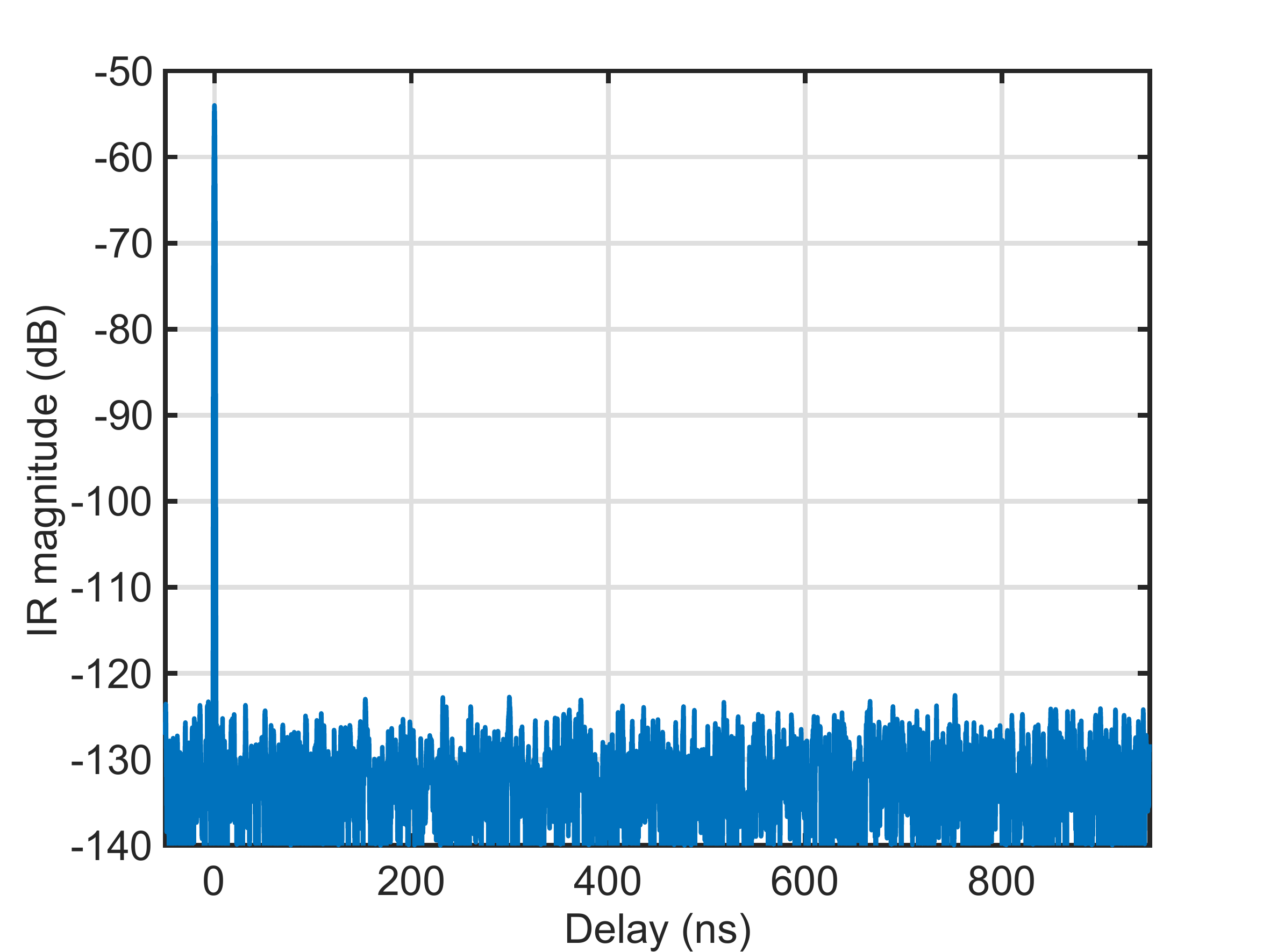}}
\caption{IR with coherent averaging over all 50,000 snapshots}
\label{fig:ultimate_cir}
\end{figure}

After applying phase tracking, all 50,000 IR snapshots were coherently averaged. The resulting IR is displayed in Fig. \ref{fig:ultimate_cir}. It still contains the main peak with a relative power of $-54$\,dB but shows a greatly reduced mean relative noise level of $-130.8$\,dB. Compared to the single IR snapshot, this calculates to a gain of $43.7$\,dB, falling only $3.3$\,dB short of the theoretical maximum gain of $47$\,dB. About $3$\,dB lost gain can be explained by noise still present in the back-to-back calibration measurement. This can be explained by the fact that the calibration measurement is also averaged over 50,000 snapshots. Thus, the calibration data used in the post-processing of the measurement data contain the same noise power as the averaged measurement data themselves. This mathematically increases the noise level in the IR by $3$\,dB. To avoid these losses, the noise in the calibration measurement must be further reduced - for example by further averaging over several calibration measurements.

Both dynamic range (DR) and maximum measurable path loss (MMPL) as defined in \cite{peter2016characterization} can directly be read off Fig. \ref{fig:ultimate_cir}. Besides the main peak with a relative power of $-54$\,dB, the IR shows the largest local maximum not related to a propagation path with a relative power of $-122.6$\,dB at a delay of $752.3$\,ns, resulting in an (instantaneous) DR of $68.6$\,dB. At the same time, the largest local maximum not related to a propagation path also directly limits the MMPL to $122.6$\,dB.

\section{Over-the-Air Measurements}
\label{sec:ota}

In addition to conducted measurements using a fixed attenuator, the channel sounder was also evaluated using over-the-air (OTA) measurements. As with the conducted measurements, 50,000 CIR snapshots were recorded and coherently averaged after phase correction. 

\begin{figure}[htb]
\begin{subfigure}{.48\textwidth}
    \centering
    \includegraphics[width=0.95\textwidth]{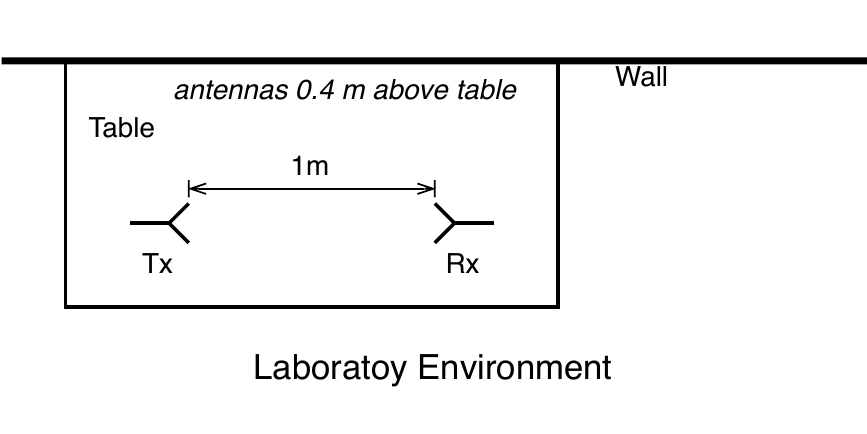}
    \vspace{-.5em}
    \caption{Open Waveguides}
    \label{fig:ota_setup_1}
\end{subfigure}

\begin{subfigure}{.48\textwidth}
    \centering
    \includegraphics[width=0.95\textwidth]{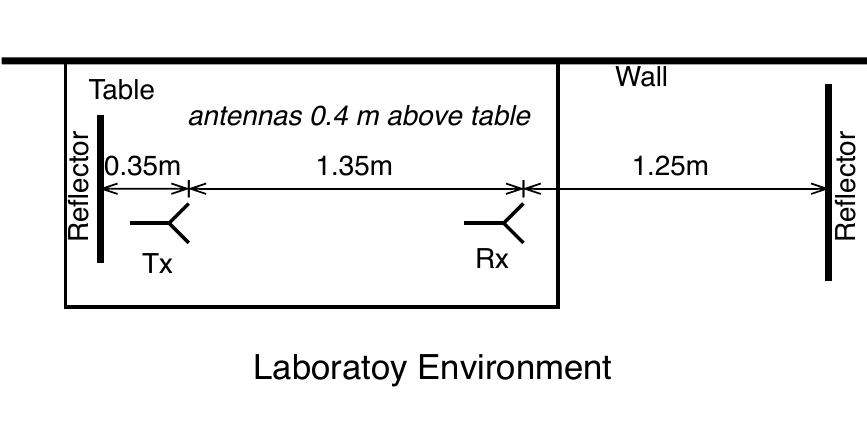}
    \vspace{-.5em}
    \caption{Horn Antennas}
    \label{fig:ota_setup_2}
\end{subfigure}
\caption{Scenarios for OTA measurements}
\label{fig:ota_setup}
\end{figure}

The first measurement was conducted using 5\,cm long open waveguides as antennas at both the transmitter and receiver, set up in way that the waveguide flanges were exactly 1 meter apart and facing each other. The setup is shown in Fig. \ref{fig:ota_setup_1}. It was expected that the relative power of the main peak in the CIR is similar to the $-72.3$\,dB measured using the vector network analyzer.

\begin{figure}[hbtp]
\centerline{\includegraphics[width=0.45\textwidth]{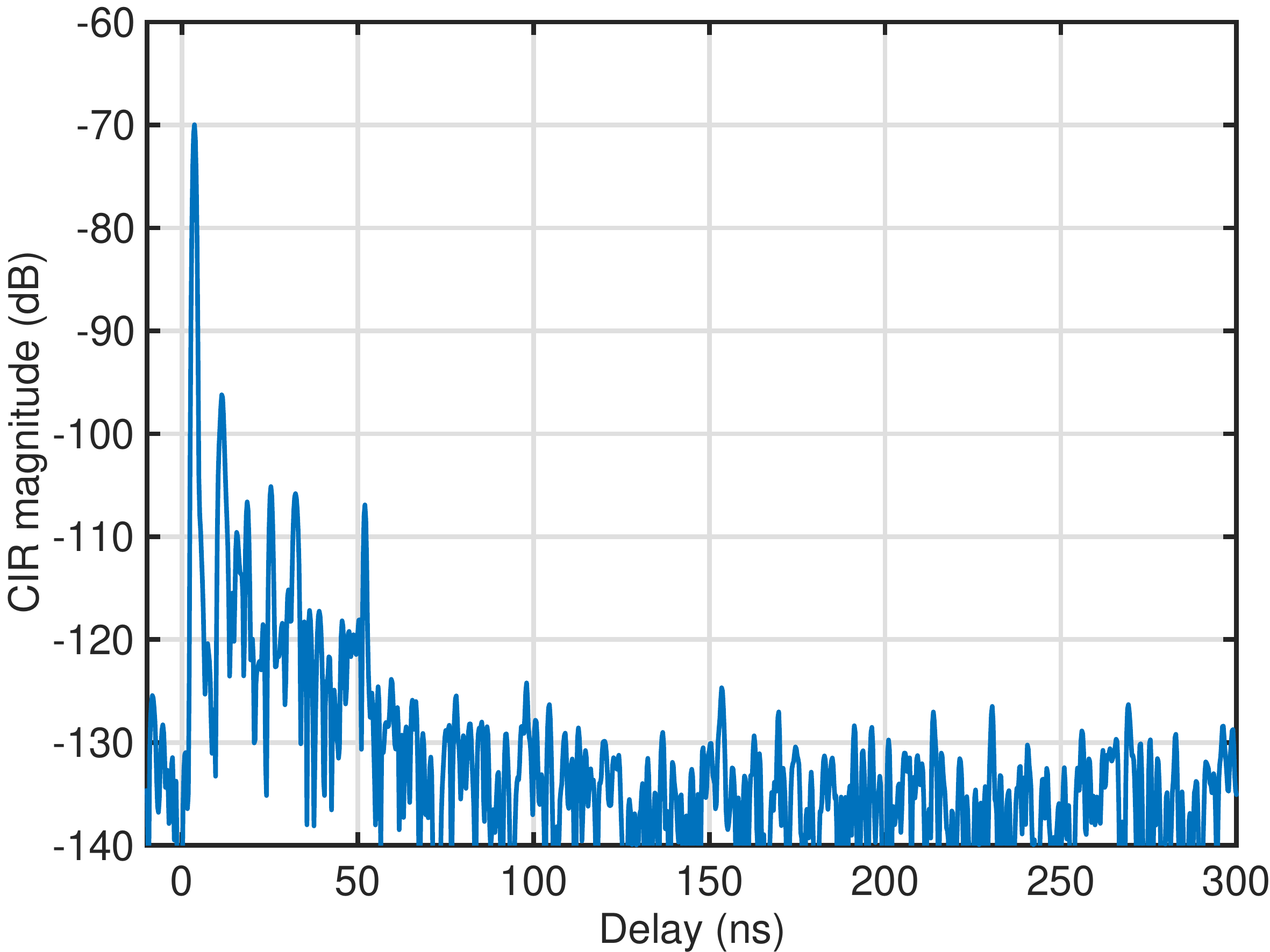}}
\caption{CIR with open waveguide antennas in d=1\,m}
\label{fig:ota_fspl}
\end{figure}

Figure \ref{fig:ota_fspl} shows the averaged CIR for the measurement with open waveguide antennas. The main peak occurs at a delay of 3.5\,ns with a relative power of $-70$\,dB, corresponding to a distance of 1.05 meters, and a mean relative noise level of $-133.7$\,dB. Considering the two 5\,cm long waveguides used for the measurements and the fact that the back-to-back calibration was done with a 5\,cm long attenuator, the measured delay and corresponding distance are extremely accurate. The relative power of the main peak is $2.3$\,dB lower than expected based on the measurements using the network analyzer. This might be due to a slight misalignment of the waveguide antennas, but gives rise to further investigations. Several multipath components are visible in the CIR at delays up to 52\,ns and with relative powers of up to $-96$\,dB. \\

\begin{figure}[htbp]
\centerline{\includegraphics[width=0.45\textwidth]{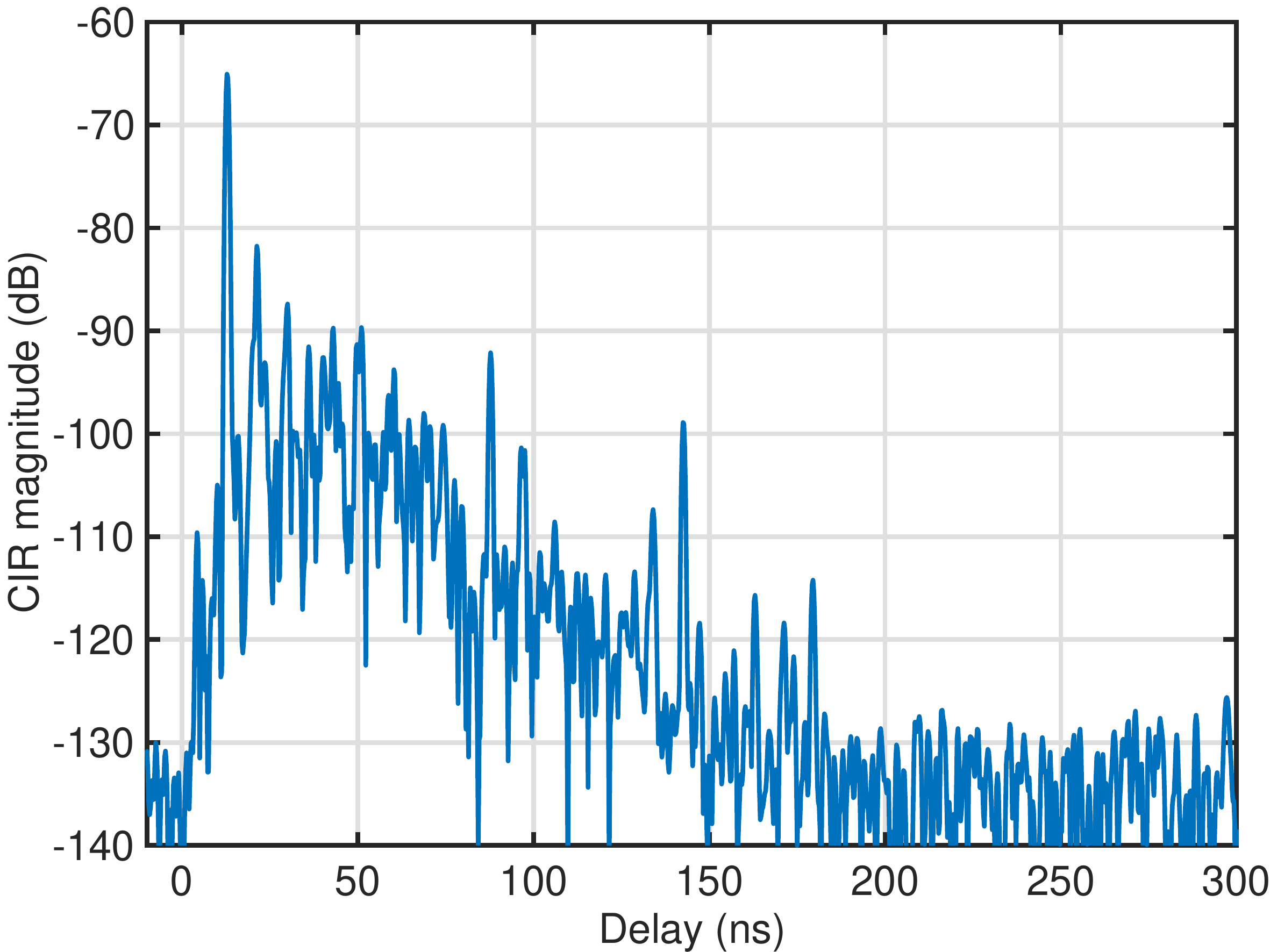}}
\caption{CIR of indoor environment with multipath components}
\label{fig:ota_mpc}
\end{figure}

The second measurement was conducted using corrugated horn antennas with a gain of 26\,dBi at transmitter and receiver, set up in line in a distance of 1.35\,m. Transmitter and receiver were facing in the same direction where a metallic reflector was set up in a distance of 1.25\,m from the receiver, as shown in Fig. \ref{fig:ota_setup_2}. An additional reflector was placed in a distance of 0.35\,m behind the transmitter. The overall propagation path between transmitter and receiver adds up to 3.85\,m. Figure \ref{fig:ota_mpc} shows the CIR averaged over all 50,000 snapshots.

\balance
The strongest multipath component arrives at a delay of 12.75\,ns with a relative power of $-65$\,dB, corresponding to a distance of 3.82\,m. This agrees well to the set up scenario. Additional strong multipath components can be seen with delays of up to 179.5\,ns.

\section{Conclusion}
\label{sec:conclusion}
In this paper, a novel, instrument-based time-domain THz channel sounder for the frequency range between 280 and 330\,GHz, developed and assembled at Fraunhofer HHI together with Fraunhofer IAF and Rohde \& Schwarz, was introduced. It offers, to the best knowledge of the authors, unprecedented performance in terms of sensitivity and dynamic range with a maximum measurable path loss of $122.6$\,dB and a dynamic range of almost $70$\,dB. This is achieved by maximizing the post-processing gain through correlation gain from the sequence and coherent averaging of a large number of sequences, made possible by phase tracking the signal in static environments.
The performance of the channel sounder was verified by over-the-air measurements in an indoor laboratory environment, which confirmed the sensitivity and highlighted the temporal resolution of the setup.

Further enhancement of the channel sounder setup could include higher measurement bandwidths and the estimation of angular information of the received signal.


\bibliographystyle{IEEEtran}
\bibliography{IEEEabrv,references}

\end{document}